\documentclass[letterpaper,10 pt, conference]{IEEEtran}
\usepackage{amsmath,amssymb}
\usepackage{array} 
\usepackage[dvips]{graphicx}
\usepackage{xspace}
\usepackage{lscape}
\usepackage{here}
\usepackage{epsfig}
\usepackage{multirow}
\usepackage{color}
\usepackage{epstopdf}

\newcommand{\beq}{\begin{equation}}
\newcommand{\eeq}{\end{equation}}

\newcommand{\barr}[2]{\begin{array}{#1}#2\end{array}}
\newcommand{\ds}{\displaystyle}
 \newcommand{\R}{\rm{I\kern-2pt R}}
\renewcommand{\theequation}{\arabic{equation}}
\newtheorem{thrm}{\bf Theorem}
\newtheorem{rmrk}{{\bf Remark}}
\newtheorem{ex}{\bf Example}
\newtheorem{crllry}{{\bf Corollary}}
\newtheorem{defi}{\bf Definition}
\newtheorem{prpstn}{{\bf Proposition}}
\newtheorem{lemme}{{\bf Lemma}}

\newenvironment{remark}{\begin{rmrk} \rm}{\end{rmrk}}

\newenvironment{definition}{\begin{defi} \rm}{\end{defi}}
\newenvironment{proposition}{\begin{prpstn} \rm}{\end{prpstn}}

\IEEEoverridecommandlockouts 

\begin{document}
\title{\vspace{4mm} Collision Avoidance and Liveness of Multi-agent Systems with CBF-based Controllers}

\author{ Mrdjan Jankovic and Mario Santillo
\thanks{M. Jankovic and M. Santillo are with
  Ford Research and Advanced Engineering,
  2101 Village Road, Dearborn, MI 48124, USA,   {\tt\small e-mail: mjankov1@ford.com, msantil3@ford.com}}%
}

\maketitle

\begin{abstract}
\noindent 
In this paper we consider multi-agent navigation with collision avoidance using Control Barrier Functions (CBF). In the case of non-communicating agents, we consider trade-offs between level of safety guarantee and liveness  -- the ability to reach destination in short time without large detours or gridlock. We compare several CBF-based driving policies against the benchmark established by the Centralized controller that requires communication. One of the policies (CCS2) being compared is new and straddles the space between policies with only local control available and a more complex Predictor-Corrector for Collision Avoidance (PCCA) policy that adjusts local copies of everyone's control actions based on observed behavior. The paper establishes feasibility for the Centralized, PCCA and CCS2 policies. Monte Carlo simulations show that decentralized,  host-only control policies lack liveness compared to the ones that use all the control inputs in calculations and that the PCCA policy performs equally well as the Centralized, even though it is decentralized.
\end{abstract}

\section{Introduction}
\noindent
With the continual advancement of automated driver assist systems, autonomous vehicles, and smarter robotic systems, management of agent-to-agent interactions has been a hot topic in recent years. These systems must be capable of automatically planning and executing their paths in real time while guaranteeing collision-free operation. These dual objectives  are often conflicting. A complicating factor in many cases arises from mixed operating scenarios, such as with multi-brand, multi-robot factories or heterogeneous driving scenarios with fully autonomous, semi autonomous, and/or human-driven agents and vehicles that both compete and cooperate. This can lead to hidden feedback loops that are only partially controllable from each agent's perspective, presenting an opportunity for the rigors of feedback control. 

In recent years, Control Barrier Functions (CBF) \cite{ames_pp,wieland} have shown great promise in providing a computationally efficient method that is both provably safe and able to handle complex scenarios (e.g. non-convex constraints). Similar to Model-Predictive Control (MPC), CBF is a model-based feedback control method that can be formulated as a quadratic program (QP) and solved online using real-time capable solvers such as  \cite{odys}. A few noteworthy differences between CBF and MPC are (i) MPC generally relies on a set of linearized dynamic systems to cover the nonlinear model range, whereas CBF deals with the nonlinear (but control affine) model directly, (ii) for non-convex constraints, MPC requires convexification, sequential convex programming, or mixed integer programming, while CBFs do not see them as such,  and (iii) MPC takes advantage of future state prediction whereas CBF does not.

In a controlled environment with all agents having the ability to communicate, a centralized controller -- an off-board computer that takes in all agents' inputs, calculates and relays the optimal action for each agent, and relies on each agent to follow this command -- may be employed \cite{wilson}. One contribution of this paper is a proof that the standard centralized CBF-based QP problem is always feasible for the distance-based barrier function.

In less controlled scenarios such as vehicles operating on roadways, or multi-brand robots operating in the same area without common communication protocols, a decentralized controller may be necessary. In this case, each agent computes and executes the best control for itself given the information that it knows (i.e. similar to how we drive vehicles). Several variations of the decentralized CBF policy exist. In particular, the Decentralized Follower \cite{borrmann} method assumes each agent takes full responsibility for collision avoidance while the Decentralized Reciprocal \cite{wang} method assigns each agent a fraction of responsibility. Robust CBF (RCBF) \cite{jankovic_aut} was used as the basis for the development of the Predictor-Corrector for Collision Avoidance (PCCA) algorithm \cite{arXiv}. We herein present a novel decentralized controller, called Complete Constraint Set (CCS2), that assigns appropriate agent responsibility and guarantees constraint adherence in the two-agent case. The PCCA and CCS2 are quazi-centralized in that they compute the best course of action for every agent with local, incomplete information. The advantage is that the corresponding QPs are always feasible.

In this paper, we present and then compare these CBF-based collision-avoidance algorithms in an interactive multi-agent setting. Specifically, after introducing each algorithm, we assess  feasibility and its effect to online algorithm implementation. We then proceed to compare algorithm metrics on liveness, collisions, and feasibility through a randomized five-agent Monte-Carlo simulation of 100 trials for each method (each set up identically for each trial). Liveness is a measure of convergence time, that is, each trial where agents successfully navigated from beginning to end locations is rated on the time they took to get there. 

The rest of this paper is organized as follows. Section \ref{sec:rcbf} reviews CBF and RCBF based control. Section \ref{sec:agents} introduces the dynamic model for the agents. Section \ref{sec:algs} reviews or introduces CBF-based controllers for collision avoidance. The simulations in Section \ref{sec:sims} consider randomized trials of five interacting agents with a stationary obstacle.\\

\noindent
{\bf Notation:} For a differentiable function $h(x)$ and a vector $f(x)$, 
$L_fh(x)$ denotes $\frac{\partial h}{\partial x} f(x)$.
A continuous function $\alpha(\cdot)$ is of class $\cal{K}$ if it is strictly increasing and satisfies $\alpha(0)=0$. We additionally assume $\alpha \in \mathcal K$ is Lipschitz continuous.

\section{Robust Control Barrier Functions reviewed}\label{sec:rcbf}
\noindent
In this section, we briefly review the concepts of Control Barrier Functions -- introduced in \cite{wieland} and later combined with quadratic programs (see, e.g. \cite{ames_pp}) -- and of robust Control Barrier Functions (R-CBF) introduced in \cite{jankovic_aut}. CBFs apply to nonlinear systems affine in the control input
\beq  \dot x = f(x) +g(x) u \label{nls} \eeq
with $x\in {\R}^n$ and $u \in {\R}^m$, while R-CBFs extend CBFs to systems with a bounded external disturbance $w(t) \in {\R^\nu}$, $ \| w(t)\| \le \bar w > 0$, of the form 
\beq \dot x= f(x) + g(x) u + p(x) w \label{dyn_w_dist} \eeq

One control objective is to regulate the system
 to the origin or suppress the disturbance (i.e. achieve input-to-state stability (ISS)) 
and we assume that there is a known baseline controller 
$u_0$ that achieves the objective. The other control objective is to
keep the state of the system in an admissible set defined by ${\mathcal C} = \{x \in {\R}^n: h(x)  \ge 0\}$
where $h(x)$ is a differentiable function. 
Here we combine definitions of CBF and RCBF into one. \\

\begin{definition} ({\em  CBF and Robust-CBF}) \
A differentiable function $h(x)$ is a CBF for the system (\ref{nls}) 
if there exists a function $\alpha_h \in \mathcal K$ such that 
\beq L_gh(x) = 0 \ \Rightarrow  \ L_fh(x) + \alpha_h(h(x)) > 0 \label{cbf} \eeq
The function $h(x)$ is an 
RCBF for the system (\ref{dyn_w_dist}) if
\beq L_gh(x) = 0 \ \Rightarrow  \ L_fh(x) - \|L_p h\| \bar w + \alpha_h(h(x)) > 0 \label{r-cbf} \eeq
\end{definition}

In the CBF case, the definition asks that, when the control over the evolution of $\dot h = L_fh +L_gh u $ is lost ($L_gh = 0$),
the rate of decrease of $h$ to 0 is not faster than $\alpha(h)$. Similarly, for the system with disturbance, the bound on the 
rate of decrease applies for the worst case disturbance. 

One advantage of CBFs for control affine systems is that they naturally lead to linear constraints on the control input $u$ 
that could be enforced online. A quadratic program (QP) is set up to enforce
the constraint, while staying as close as possible to the baseline (performance) control input $u_0$:  \\

\noindent
{\bf (R)CBF QP Problem}: Find the control $u$ 
 that satisfies
\beq \barr{l}{\ds \min_u \|u-u_0\|^2 \ \  {\rm subject \  to}  \\*[2mm]
 \ds F_i \ge 0, \ \  i = 0, 1,\  {\rm or}\  2 }\label{rQP} \eeq
where we select $F_0 = L_fh(x) + L_g h(x) u + \alpha_h(h(x)) $ if $h$ is a CBF for the system (\ref{nls});
$F_1 = L_fh(x) - \|L_p h(x)\| \bar w + L_g h(x) u + \alpha_h(h(x))$ if $h$ is an RCBF for the system (\ref{dyn_w_dist}) with an unknown disturbance; or $F_2=
L_fh(x) + L_p h(x) \hat w + L_g h(x) u + \alpha_h(h(x))$  when an estimate/measurement
of the disturbance is available. \\

The available results (e.g. \cite{ames_pp, jankovic_aut}) guarantee that the resulting control is Lipschitz continuous, the barrier constraint $F_i$ 
is satisfied, which implies that $h(x(t)) \ge 0, \forall t$ and the admissible set $\mathcal C$ is forward invariant. Note that the strict ``$>$'' is needed in the (R)CBF definition (\ref{cbf}) and (\ref{r-cbf}) to guarantee
Lipscitz continuity of the control law \cite{jankovic_aut}, or,  alternatively, $L_gh(x) \not = 0, \forall x \in {\mathcal C}$ needs to be assumed as in \cite{ames_pp}. 

In the barrier functions considered in the rest of the paper, the control input does not 
appear in their first derivative. So the standard definition of CBF and R-CBF offered above does not apply. The inputs appear in the second derivative of $h$, so we follow the 
ideas of \cite{nguyen, xu_rel_deg} for dealing with higher relative degree barrier functions.  
Instead of enforcing $\dot h + \alpha_h(h) \ge0$, we switch to linear barrier dynamics and enforce
\beq \ddot h + l_1 \dot h + l_0 h\ge 0 \label{r2_barrier}\eeq
as the QP constraint. The parameters
$l_0, l_1$ should be selected so that the two roots $\{-\lambda_1, -\lambda_2\}$   of the polynomial
$s^2+l_1 s + l_0 = 0$ are negative real ($\lambda_{1/2} = \frac{l_1\pm\sqrt{l_1^2 -4l_0}}{2}$ ) 
and the $h$ dynamics in (\ref{r2_barrier}) is stable.
It is a matter of straight-forward calculation to show that, if the barrier constraint (\ref{r2_barrier}) holds,  the set 
${\mathcal C}^* = \{ (x): h(x) \ge 0, h(x) \ge -\frac{1}{\lambda_i }\dot h(x) \}$, where $-\lambda_i$ is either of the two eigenvalues,
is forward invariant. With ${\mathcal C}^*\subset {\mathcal C}$ the original constraint $h(x) \ge 0$ will be satisfied.
 
For the second-order barrier, the QP constraints that need to be enforced for ${\mathcal C^*}$ to be forward invariant are 
\beq  F_0 = L_f^2 h + L_gL_f h u + l_1 L_fh + l_0 h \ge 0 \label{r2-constr} \eeq
in the case of the CBF for the system without disturbance;
\beq F_1 = L_f^2 h - \|L_pL_fh \|\bar w + L_gL_f h u + l_1 L_fh + l_0 h \ge 0 \label{r2_const_barw} \eeq
for an RCBF with unknown disturbance bounded by $\bar w$;  or
\beq F_2 = L_f^2 h + L_fL_p h \hat w + L_gL_f h u + l_1 L_fh + l_0 h \ge 0 \label{r2_const_hatw} \eeq
for an RCBF with known disturbance estimate $\hat w$.

\section{Holonomic Agent Model}\label{sec:agents}
\noindent
In the literature, agents are typically modeled either with holonomic double integrators in the X-Y plane or as 
as  non-holonomic ``unicycle'' or ``bicycle'' model. For simplicity, here we consider only the first option.
An agent is modeled as a circle of radius $r_0$ with the center motion
given by  the double integrator in each dimension:
\beq \barr{l} {\dot x= v_x \\
         	     \dot y = v_y \\
	     	      \dot v_x = u_x, \\
		      \dot v_y = u_y} \label{agent_i} \eeq

The relative motion between any two agents $i$ and $j$ is given by
\beq \barr{l}{ \dot \xi_{ij}= v_{ij} \\ 
\dot v_{ij} = u_i - u_j} \label{diff_motion} \eeq
where the indices correspond to agents $i$ and $j$, $\xi_{ij} = [x_i - x_j, y_i - y_j]^T$ is the center-to-center (vector)
displacement between the two agents and $v_{ij} = [v_{xi} - v_{xj}, v_{yi} - v_{yj}]^T$ 
is their relative velocity.
Our goal is to keep the $\|\xi_{ij}\|$ larger than $r \ge 2r_0$ (with the  distance $r$ strictly greater than $2 r_0$ we are providing a ``radius margin" relying on
robustness of barrier functions (see \cite{xu}) to push the states out of the inadmissible set).  
To this end, we define a barrier function
\beq h(\xi_{ij}) = \xi_{ij}^T\xi_{ij} - r^2  \label{h_ij} \eeq
with the goal to keep  it greater than 0. The advantages of this barrier function over alternative ones used for multi-agent collision avoidance 
is that (i) we can prove the feasibility of the Centralized and a few decentralized QPs and (ii) it allows
a radius (barrier) margin because the calculation does not collapse when $h(\xi_{ij} )<0$. 
A disadvantage  of the barrier function is that is has relative degree two from all 4 inputs. Because of this, 
we apply the approach described in Section \ref{sec:rcbf} and form a CBF barrier constraint:
\beq  F_{ij} := \ddot h + l_1 \dot h + l_0 h = a_{ij} + b_{ij}(u_i - u _j )\ge 0  \label{h_constr} \eeq
where $a_{ij} = 2v_{ij}^Tv_{ij}+ 2 l_1 \xi_{ij}^T v_{ij} +  l_0 (\xi_{ij} ^T\xi_{ij} - r^2)$, $b_{ij}  = 2 \xi_{ij} ^T$, 
and $u_i$ and $u_j$ are the control actions of the two
agents. The function $h$ is a CBF for the system (\ref{diff_motion}) because $L_gh= 2b_{ij} \not = 0$ unless the two agents completely overlap
(that is, are well past the point of collision). As a result, we can always enforce positive invariance of the 
the admissible set ${\mathcal C}^*_{ij}= \{ (\xi_{ij}, v_{ij}): h(\xi_{ij}) \ge 0, h(\xi_{ij}) \ge -\frac{1}{\lambda_1 }\dot h(\xi_{ij}, v_{ij}) \}$ with $-\lambda_1$ as one of the two eigenvalues as discussed above. 

As we shall see below, with the Centralized controller, we can guarantee that the agents are not colliding even if there is no barrier margin:
$r = 2r_0$. In the non-ideal case -- for example, discrete-time implementation, and this is the only way a QP could be implemented --
some very small barrier function violations start to appear. Non-centralized controllers might have larger violations because
two agents $i$ and $j$ compute $u_i$ and $u_j$ independently, based on different information available to them. 
To the extent they don't agree, a difference between the left hand sides of the centralized barrier constraint $F_{ij}^c \ge 0$, which guarantees collision avoidance, 
 and the actual $F_{ij}^a \ge 0$ resulting from independent control calculations could appear. Let's denote this quantity
by $\Delta F_{ij}$. If $l_0(r^2 - r_0^2) \ge \Delta F_{ij}$ when the two agents are in close proximity, then
the constraint with $h_0= \|\xi_{ij}\|^2 -2r_0$ as the barrier function will be satisfied and $h_0$ would remain positive for all $t$.

\section{CBF-Based Collision Avoidance Algorithms}\label{sec:algs}
\noindent
In the absence of other agents, we assume that each agent has its own preferred control action $u_{0i}$ (for agent $i$), 
computed independently of the collision avoidance algorithm. In this paper, we have assumed that the final destination for
each agent is known to the agent and used the Linear Quadratic Regulator (LQR) controller to compute $u_{0i}$'s. 
With the full knowledge of $u_{0i}$'s and with the barrier constraints between each two agents defined as in the previous section, 
the Centralized controller could be set up as the solution to the following quadratic program:\\

\noindent
{\bf Centralized QP}: Find the controls $u_i, i=1,\ldots, N_a$ 
\beq \barr{l}{\ds \min_{u_{1}, \ldots u_{N_a} } \sum_{i=1}^{N_a} \|u_i -u_{0i}\|^2 \ \  {\rm subject \  to}  \\*[2mm]
 \ds a_{ij} + b_{ij}(u_i - u_j )\ge 0  \ \ \forall i,j =1,\ldots, N_a , i\not = j}\label{rQPc} \eeq
where $a_{ij}$ and $b_{ij}$ are defined in the previous section and $N_a$ is the number of agents. 

The QP solution could be computed by a central node and communicated to the agents, or each agent could solve 
the QP independently, which still requires communication between them. If this QP problem is feasible, and this is proven below, the control action would satisfy all the  barrier constraints (\ref{h_constr}) and guarantee collision-free operation (see  \cite{wang}).

Without communication, the base control action $u_{0j}$ for the target (i.e. other) agents are not available to the host  $i$ (the agent doing the
computation). In this case, each agent could implement an on-board decentralized controller.  One version, included here because it
resembles many defensive driving policies, is for each agent to accept full responsibility for 
avoiding all the other agents. Borrowing nomenclature from  
Game Theory, we refer to this policy as ``Decentralized Follower'' (DF): \\

\begin{figure}[htbp!]\vspace{-.05in}
    \centering
\includegraphics[scale=0.40]{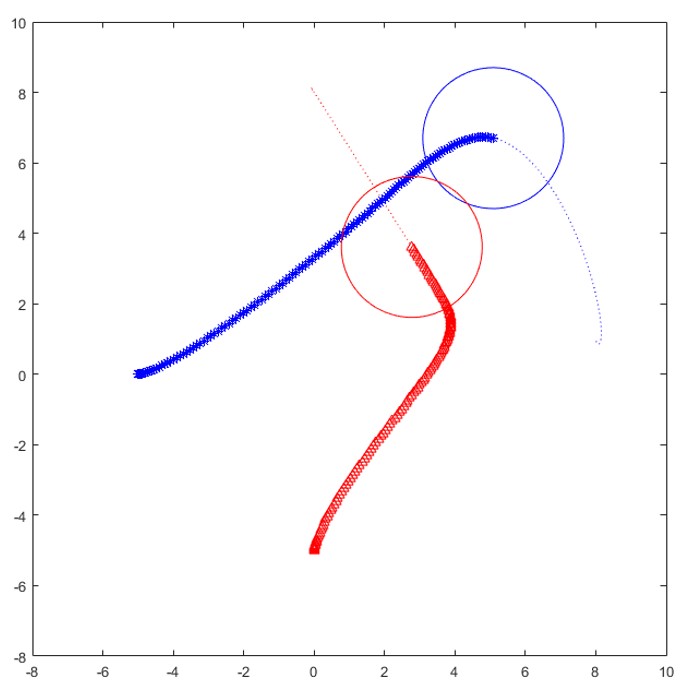}
    \vspace{-0.1in}
    \caption{Two agents colliding while crossing paths with the Decentralized Follower policy.}
    \label{fig:DFA}
\end{figure}

\noindent
{\bf Decentralized Follower QP} (for agent $i$): Find the control $u_i$ for the agent $i$
 that satisfies
\beq \barr{l}{\ds \min_{u_{i}}  \|u_i -u_{0i}\|^2 \ \  {\rm subject \  to}  \\*[2mm]
 \ds a_{ij} + b_{ij}u_i\ge 0  \ \forall  j =1,\ldots, N_a , j\not = i}\label{rQPd} \eeq 
This formulation is essentially the same as in \cite{borrmann} but a different barrier function is used, as described above.
The agent $i$ has only its own actions ($x$ and $y$ accelerations) to avoid all other agents and 
there are no guarantees that the DF QP is feasible. Even when it is feasible and the agents apply the same defensive algorithm, 
there are no collision avoidance guarantees. The reason is that each agent knows only its own acceleration $u_{0i}$, 
and may assess it safe to apply. That is, if $a_{ij}+b_{ij}u_{0i} \ge 0$, $\forall j$, agent $i$ would consider $u_{0i}$ safe to apply.
Similarly, agent $j$ might find that $u_{0j}$ is safe to apply. However, $a_{ij}+b_{ij}u_{0i} \ge 0$
and $a_{ij} - b_{ij}u_{0j} \ge 0$ does not imply $a_{ij}+b_{ij} (u_{0i} - u_{0j}) \ge 0$, which would actually guarantee
collision avoidance. Indeed, our simulations show that, even with only two agents, there could be a collision as seen 
by overlapping circles in Figure \ref{fig:DFA}. Each agent only implements the (identically tuned) navigation policy described by (\ref{rQPd})
with $u_0$'s coming from an LQR controller.
We note that (i) the collision did not happen during braking (i.e. when the agents are
approaching one another), but when they both optimistically assume it is safe to accelerate; (ii) even though the algorithm 
updates controls every 50ms with new position and velocity information, the collision is not avoided when 
the constraint is violated. 

\begin{figure}[htbp!]\vspace{-.1in}
    \centering
\includegraphics[scale=0.37]{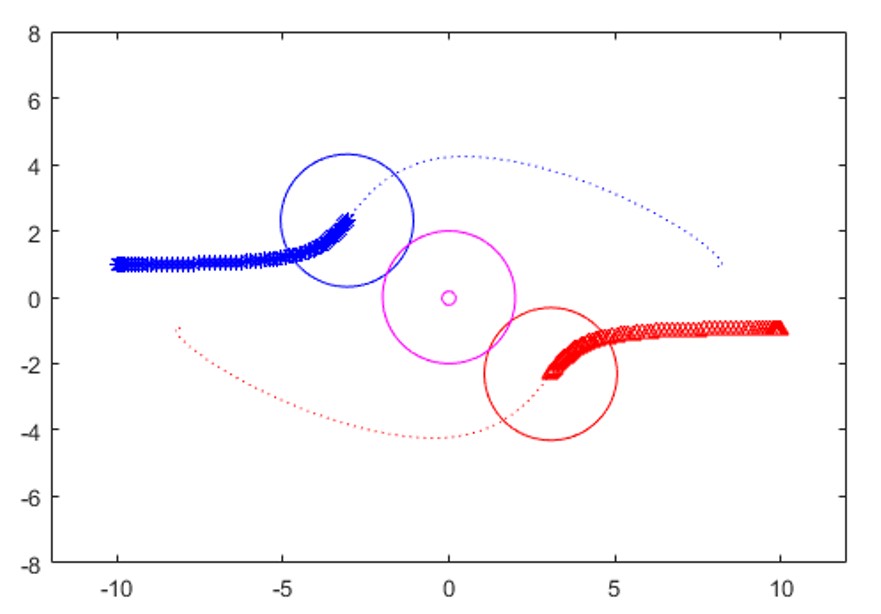}
    \vspace{-0.1in}
    \caption{Two agents passing a stationary one from opposite directions, all running the Decentralized Reciprocal policy.}
    \label{fig:DRA}
\end{figure}

To  improve the DF performance, the  ``Decentralized Reciprocal" (DR) policy was introduced in \cite{wang}:\\
\noindent
{\bf Decentralized Reciprocal QP} (for agent $i$): Find the control $u_i$ for the agent $i$
 that satisfies
\beq \barr{l}{\ds \min_{u_{i}}  \|u_i -u_{0i}\|^2 \ \  {\rm subject \  to}  \\*[2mm]
 \frac{1}{2} a_{ij} + b_{ij}u_i\ge 0  \ \  \forall  j =1,\ldots, N_a , j\not = i}\label{rQPr} \eeq 
 The only difference from the DF version is the $\frac{1}{2}$ factor multiplying $a_{ij}$, meaning that
 each agent assumes half the responsibility for avoiding collision (we assumed all the agents are the same). 
 The method was shown in \cite{wang}
 to guarantee constraint adherence and, hence,  collision avoidance as long as it is feasible and, when it is not, 
proposed a braking action. Braking, however works only if all agents, even those with 
 feasible QP, apply it at the same time. To illustrate the issue, we consider three agents, with two passing the 
 stationary one in the middle as shown in Figure \ref{fig:DRA}. The DR QP turns out infeasible for the agent in the middle, 
 but, because it was already stationary, the braking applied has no effect. The other two have feasible QP's and keep applying the solutions.
 This leads to collisions as shown in Figure \ref{fig:DRA} because the expected
 half contribution towards avoiding collisions by the agent in the middle has not been met.
 One possible approach to avoid the problem is for each agent to run DR-QP's for itself and each of the other agents and
 brake when one of the QPs turns infeasible (no need to know $u_{0j}$'s to assess feasibility). This would increase 
 computational footprint and raise the issue of when to stop braking: as soon as all QP's become feasible, or
 only after all agents have stopped. The former might lead to a jerky motion, while the latter would suffer from reduced liveness. 
 In the simulation section, we have allowed the QP solver to resolve the feasibility issue by selecting control with the 
 smallest constraint violation counting on eventual application of the radius margin.

Instead of $N_a$ DR-QP's being solved by the host to assess feasibility for all the other agents, one could consider setting up a
single QP that includes all the constraints. The problem, of course,  is that the host does not know other agents' preferred accelerations, so 
zeros are used instead of unknown $u_{0j}$'s. We refer to 
this policy as the ``Complete Constraint Set 2" (CCS2), with the
meaning of ``2" in the acronym explained below: \\

\noindent
{\bf CCS2 QP} (for agent $i$): Find control actions $u_{ij}, j=1,\ldots, N_a$ 
such that
\beq \barr{l}{\ds \min_{u_{i1}, \ldots, u_{iN_a}}  \sum_{j=1}^{N_a} \|u_{ij}\|^2 \  {\rm subject \  to}  \\*[2mm]
 \ds a_{ij} + 2 b_{ij} u_{0i} + b_{ij}(u_{ii}- u_{ij})\ge 0  \ \forall  j =1,\ldots, N_a , j\not = i \\*[2mm]
 \ds a_{jk} + b_{jk}(u_{ij} - u_{ik} )\ge 0  \ \ \forall  j,k =1,\ldots, N_a \\*[2mm]
 \ds \hspace*{2.0in} j\not = k \ {\rm and} \ j,k \not = i}\label{rQPccs} \eeq
The agent $i$ uses CCS2 policy to find the control action for all the agents and implements its own: $u_i = u_{ii}^* + u_{0i} $ where
$u_{ii}^*$ denotes the solution to the QP problem.
The CCS2 QP is guaranteed to be feasible
(see Proposition \ref{prop_feasibility} below).

To understand the multiplier ``2" in front of $b_{ij} u_{0i}$, first note that with ``$1$'' in its place it would be 
equivalent to the centralized QP (\ref{rQPc}) with all $u_{0j}$, except $j = i$, (i.e. all the unknown ones) set to 0
 and with a variable change for $u_{ii}$. While the complete set of
 constraints guarantees that agents will correctly split the responsibility for $a_{ij}$, other agents do not know about $u_{0i}$ so 
 the responsibility could not be split. The multiplier 2 works in the case of two agents implementing  the CCS2, when 
 constraint adherence could be established using the closed form solution from \cite{arXiv}.
In the multi-agent case, this is not the case -- one could find situations when the constraints will not be satisfied. Still, because it is always feasible and,
as it turned out, more lively than the two decentralized policies, we have included it in the comparison. 

An idea to use RCBF's to robustify the CCS1 (apologies for the abuse of notation) against the missing information 
led to the development of the ``Predictor-Corrector for Collision Avoidance" (PCCA) method (see \cite{arXiv}):\\

\noindent
{\bf PCCA QP} (for agent $i$): Find control actions $u_{ij}, j=1,\ldots, N_a$ 
such that
\beq \barr{l}{\ds \min_{u_{i1}, \ldots, u_{iN_a}} \left(  \|u_{ii} -u_{0i}\|^2 +\sum_{j=1, j\not = i}^{N_a} \|u_{ij}\|^2 \right)\  {\rm subject \  to}  \\*[2mm]
 \ds a_{ij} + b_{ij}(u_{ii}- u_{ij}- \hat w_j)\ge 0  \ \forall  j =1,\ldots, N_a , j\not = i \\*[2mm]
 \ds a_{jk} + b_{jk}(u_{ij}+ \hat w_j - u_{ik} - \hat w_k)\ge 0  \ \ \forall  j,k =1,\ldots, N_a \\*[2mm]
 \ds \hspace*{2.0in} j\not = k \ {\rm and} \ j,k \not = i}\label{rQPp} \eeq
and implement its own: $u_i = u_{ii}^*$.

This setup resembles the CCS2 (\ref{rQPccs}) with all the agent-to-agent constraints accounted for, but with no ``$\times 2$"
multiplier applied. Instead, the (fictitious) disturbance terms $w_{ij}$  have been added to the $u_{ij}$ ($i \not = j$). They represent the
uncertainty of agent $i$'s computation of agent's $j$ acceleration. One could put an upper limit on this uncertainty and
proceed with RCBF using the worst case disturbance. Instead, the PCCA uses the estimated disturbance as a difference
between the control action for agent $j$ ($u^*_{ij} $) computed by the host (agent 
$i$) with the action agent $j$ actually implemented ($u_j$):
\beq \hat w_j = u_j - u^*_{ij} \label {hatwj} \eeq 
Because $u_{ij}^*$ requires knowing $\hat w_j$ and vice versa, a (static) feedback loop is created.
To break this static loop, one could use either the value from the previous sample (the controller solving
the QP could only be implemented in discrete time) or a low pass filter.

The paper \cite{arXiv} considered the case of two agents and, with discrete single sample delay, proved that the error in
enforcing the constraint is of the order of the sample time $\Delta T$ -- the smaller the sampling time, the smaller the error. 
Moreover, \cite{arXiv} showed that, even if one agent is not cooperating, 
the other agent takes over full responsibility for collision avoidance also producing the error of the order of the sample 
time. In contrast to CCS2, we could not find a multi-agent case 
when the real constraints would not be satisfied (within $\Delta T$ accuracy) but the general proof is not available. We also note that PCCA assumes 
information (measurement) of other agents' acceleration.

We now show that both the Centralized, CCS2, and PCCA QP's are feasible and,
to the best of our knowledge, this is a new result. 
The problem is nontrivial because there is a scenario
where we have more active QP constraints that the linearly independent (row) vectors
multiplying control inputs. 
This situation also prevents the standard approach to establishing Lipschitz continuity of optimal programs from being used \cite{morris}
(note: we are not implying the controller is not Lipschitz continuous). Instead, we 
offer the following feasibility result and associated continuity: \\
\begin{proposition} \label{prop_feasibility} The Centralized QP (\ref{rQPc}), CCS2 QP (\ref{rQPccs}), and PCCA QP (\ref{rQPp}) are always feasible in the admissible set 
${\mathcal C}^* = \{ x \in {\R}^n: h_{ij}(x) \ge 0, h_{ij}(x) \ge -\frac{1}{\lambda_1 }\dot h_{ij}(x), i,j \in\{1, \ldots, N_a\}, i\not = j \}$,
the solution in each case is unique and the resulting optimal control law $u^*$ is a continuous function of $(u_0, \xi,v)$. 
\end{proposition}

\noindent
{\bf Proof}: \ 
Consider the Centralized policy constraint $F_{ij} := a_{ij} + b_{ij}(u_{i} - u_j)$. Using the definition of $a_{ij}$ and $b_{ij}$ we obtain
\[ F_{ij} \ge 2\|v_{ij}\|^2 + 2\xi_{ij}^T(u_i-u_j) + 2\lambda_1\xi_{ij}^T v_{ij} \]
where the last term is obtained by using
$h_{ij}\ge \frac{-1}{\lambda_1 }\dot h_{ij}$ (from the definition of ${\mathcal C}^*$) and
$l_1 - l_0/\lambda_1 = \lambda_1$.
From here, we proceed to construct a feasible $u$ by selecting one that satisfies 
\beq u_i - u_j = -\lambda_1 v_{ij}  \label{u12_ind} \eeq
which results in $F_{ij} \ge 2\|v_{ij}\|^2 \ge 0$. 

We proceed by using mathematical induction. 
For the first two agents, we pick any $u_1$ and $u_2$ that satisfy
(\ref{u12_ind}) 
where, in this case, $i = 1$ and $j=2$.
For example, we could select $u_1 = 0, u_2 = \lambda_1 v_{12}$.
Now, assume that for the first $l-1$ agents we have selected 
control inputs $u_1, \ldots, u_{l-1}$ such that the condition (\ref{u12_ind}) 
 holds for all $i,j = 1,\ldots, l-1, \ i\not = j$. Adding the $l$-th agent
 we first consider $F_{1l} \ge 2\|v_{1l}\|^2 + 2\xi_{1l}^T(u_1-u_l +\lambda_1 v_{1l})  $.
We select $u_l = u_1 + \lambda_1 v_{1l}$. This makes $F_{1l}\ge 0$ and we need to show that 
 all the other constraints are satisfied. Because $v_{il} = v_{ij} + v_{jl}$, $v_{ij} = -v_{ji}$, and 
 (\ref{u12_ind}) holds for $i,j = 1,\ldots, l-1, \ i\not = j$ by the induction assumption, 
 for all $i = 2, \ldots, l-1$ we have 
 \[ \barr{l}{ \lambda_1 v_{il} + u_i - u_l = \lambda_1 (v_{i1} + v_{1l}) + u_i - u_l \\*[2mm]
  \hspace*{6mm} = -u_i+u_1 + \lambda_1 v_{1l} + u_i - u_l = 0 }\]
Thus, for all $i = 1, \ldots, l-1$, $F_{il} \ge 2\|v_{il}\|^2 \ge 0$ and the induction argument completes the feasibility part for the 
Centralized QP.  Because the optimal program has quadratic cost and linear constraints, there is a single solution that 
 is continuous  (see \cite{robinson_day}) in the parameters ($u_0$, $\xi$, $v$). 
 
 Feasibility of CSS2 follows because, by changing the variables, the constraint set takes the same form as that of the Centralized Controller
 with only the cost function being different. The same applies for PCCA.
\hspace*{\fill} $\bigtriangledown$ \\
 
\begin{remark} 
From the proof of Proposition \ref{prop_feasibility} it is clear that we have one extra degree of freedom assuring feasibility 
even if one agent, say agent 1, is non-interacting, but with known acceleration. Second,  if we introduce a fictitious, stationary  agent 1, 
a feasible action is if all other agents apply
$u_i = -\lambda_1 v_i$ -- a braking policy where deceleration is proportional to the 
agents velocity. 
 Note that  the proportional braking policy is a suboptimal option
(it leaves $F_{ij} \ge 2\|v_{ij}\|^2$) proving feasibility, not an external action to be applied when the respective QP is not feasible.
The above consideration also shows that the deceleration for each agent need not be larger than $ \lambda_1 v_i$,
which provides feasibility even when the deceleration is limited, but the agent speed would have to be limited too. 
\end{remark}

\section{Simulation Results}\label{sec:sims}
\noindent
We now compare the CBF collision-avoidance algorithms reviewed above by Monte-Carlo simulation of five agents maneuvering in an enclosed area. In all cases, the agents are modeled as a circle of radius $r_0 = 2$ with the center motion given by a double integrator in two dimensions as in \eqref{agent_i}. A static outer circle of radius $R_0 = 11$ acts as an additional (soft) barrier constraint to enclose the space containing all five agents. The controller sample time is chosen to be $\Delta T = $50 ms, and the baseline controller $u_{0i}$ for each agent is computed by LQR with $Q=0.2I_4$ and $R=I_2$. For computation of the QP constraints \eqref{h_constr}, we choose $l_0 = 6$ and $l_1 = 5$ to satisfy $l_1^2 \ge 4l_0$. 
All the algorithms use this same set of parameters. 

\begin{figure}[t!]\vspace{-.0in}
    \centering
\includegraphics[scale=0.78]{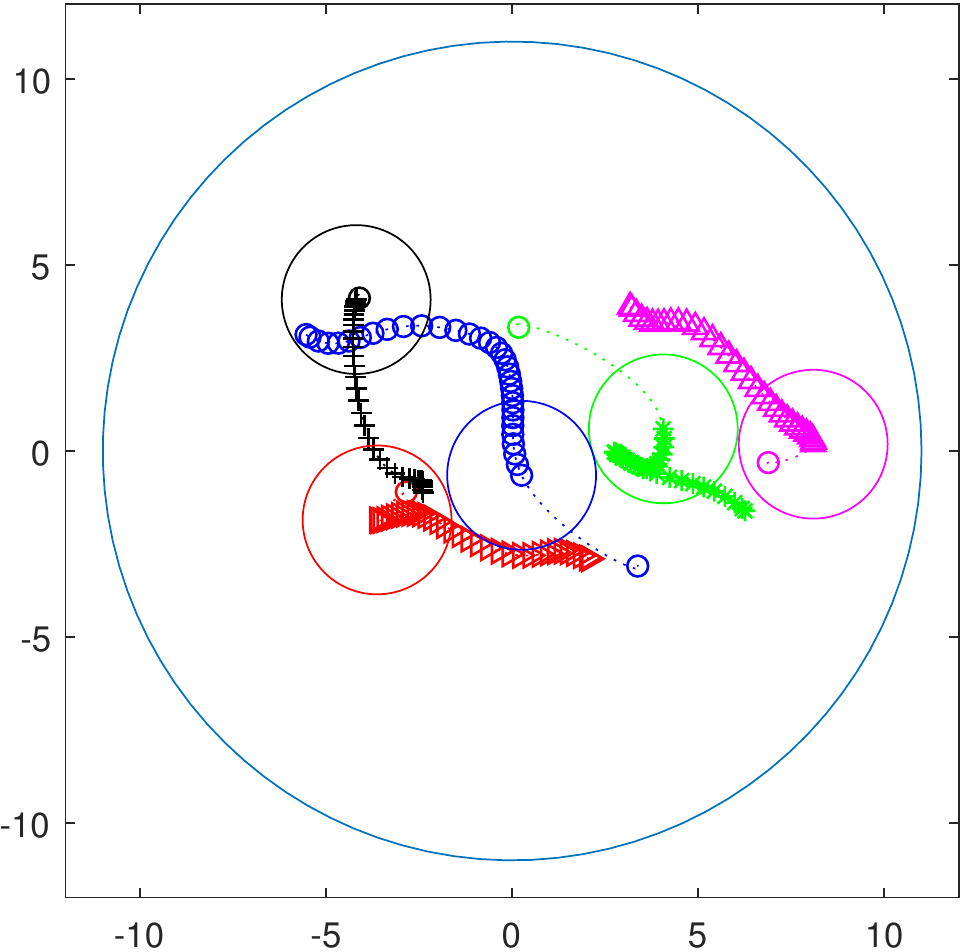}
    \vspace{-0.1in}
    \caption{5-agent simulation time snapshot depicting past and future paths. Each agent has a randomly defined non-conflicting beginning/end location.}
    \label{fig:5agent}
\end{figure}

To randomize each simulation trial, each agent is initialized with a random beginning and end location somewhere within the outer static circle's area. These locations are then assessed for any agent-to-agent overlap as well as overlap between each agent and the outer static circle. If any physical overlap is calculated, new beginning and end locations are assigned until feasible initial and final conditions are ensured. Each of the algorithms then ran from the same 100 feasible initial positions to the corresponding final positions. Figure \ref{fig:5agent} shows a time snapshot of one of these randomized simulation trials -- the agents, their beginning and end locations, as well as their past and future paths are all displayed.

To accurately compare the algorithms, a set of metrics was devised to compare liveness, collisions, and feasibility. Liveness is a measure of convergence time; for the purpose of this analysis, we assess how long it takes for all the agents to reach within a position error of 0.1 units from their destinations as well as have velocity magnitude less than 0.1 units/sec. Each simulation was run for 100 seconds and assessed for convergence. It was found that all non-convergent runs at 100 seconds had gridlocked and were not expected to converge.
We did not use the deconfliction algorithms for gridlocks because they need a preferred passing direction to be agreed up front \cite{wang} or determined on line, which
assumed agent-to-agent communication in \cite{celi}.

\begin{figure}[htbp!]
    \centering
\includegraphics[scale=0.66]{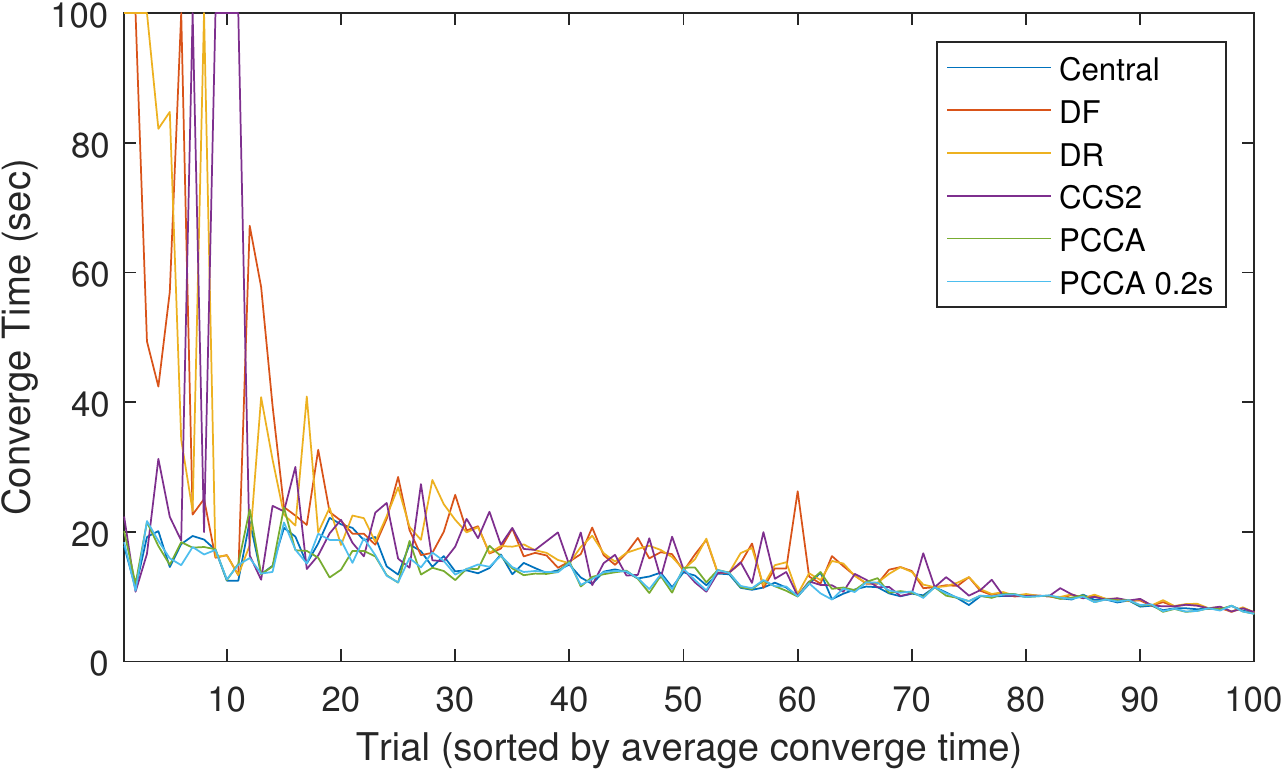}
    \vspace{-0.1in}
    \caption{Convergence time for 100 Monte Carlo simulation runs, sorted by average from max to min excluding non-convergent trials.}
    \label{fig:time}
\end{figure}

The results shown in Table \ref{tab:compare} depict the aggregated results from 100 Monte-Carlo simulations for the Centralized \eqref{rQPc}, DF \eqref{rQPd}, DR \eqref{rQPr}, CCS2 \eqref{rQPccs}, and PCCA \eqref{rQPp} policies without any additional radius margin added. PCCA was implemented with either a sample delay ($\Delta T = 50$ms) or a low-pass filter with a time constant of 0.2 sec to break the algebraic loop between (\ref{rQPp}) and \eqref{hatwj}. Looking at Table \ref{tab:compare}, we see the minimum convergence time over all 100 simulations is similar for each algorithm,
%
%
\begin{table}[b!]\vspace{-0.1in}
	\caption{Metrics for CBF-based collision-avoidance algorithms from 100 Monte-Carlo simulation runs with no radius margin}	
	\begin{tabular}{l|ccc|c|c|c|}
	\cline{2-7}
	                                & \multicolumn{3}{c|}{Converge Time (sec)}             &                & \# no    & \#         \\
	                                & min                   & max                   & mean & $h_{\rm{min}}$ & converge & infeasible \\ \hline
	\multicolumn{1}{|l|}{Central}      & \multicolumn{1}{c|}{7.45} & \multicolumn{1}{c|}{22.15} & 12.98    & -0.002    & 0         & 0          \\ \hline
	\multicolumn{1}{|l|}{DF}           & \multicolumn{1}{c|}{7.55} & \multicolumn{1}{c|}{67.20} & 17.44    & -2.84      & 3         & 27        \\ \hline
	\multicolumn{1}{|l|}{DR}           & \multicolumn{1}{c|}{7.55} & \multicolumn{1}{c|}{84.75} & 17.26    & -1.53      & 4         & 32        \\ \hline
	\multicolumn{1}{|l|}{CCS2}         & \multicolumn{1}{c|}{7.60} & \multicolumn{1}{c|}{31.25} & 14.63    & -1.35      & 4         & 0          \\ \hline
	\multicolumn{1}{|l|}{PCCA}         & \multicolumn{1}{c|}{7.35} & \multicolumn{1}{c|}{23.75} & 12.76    & -0.015      & 0         & 0          \\ \hline
	\multicolumn{1}{|l|}{PCCA$_{0.2}$} & \multicolumn{1}{c|}{7.35} & \multicolumn{1}{c|}{21.65} & 12.68    & -0.067    & 0         & 0          \\ \hline
	\end{tabular}
	\vspace{-0.1in}
	\label{tab:compare}
\end{table}
%
though the maximum is quite varied. Additionally, the max and mean values do not include the non-convergent simulation results for DF, DR, and CCS2. Both the DF and DR algorithms exhibit less liveness and generally take longer for all agents to converge as previously reported in the literature (e.g. \cite{wang}). We see this better in Figure \ref{fig:time}, where the simulations trials are sorted by the average convergence time for each trial over all algorithms from maximum to minimum. Generally, the convergence times and liveness of the Centralized and PCCA algorithms are similar, while DF, DR, and CCS2 exhibit longer times to converge, if at all.

%

It has been established above that the Centralized, CCS2, and PCCA controllers are always feasible and the simulations confirmed this. 
However, nearly a third of the DF and DR simulations exhibited infeasible QPs at some point. In this case, the QP solver \cite{odys} was configured to return the "least infeasible" solution before implementing the control. Except for slacked constraints on the outer static circle, the algorithms were implemented in pure form without alternative actions to handle infeasibility.

\begin{figure}[htbp!]\vspace{-.05in}
    \centering
\includegraphics[scale=0.66]{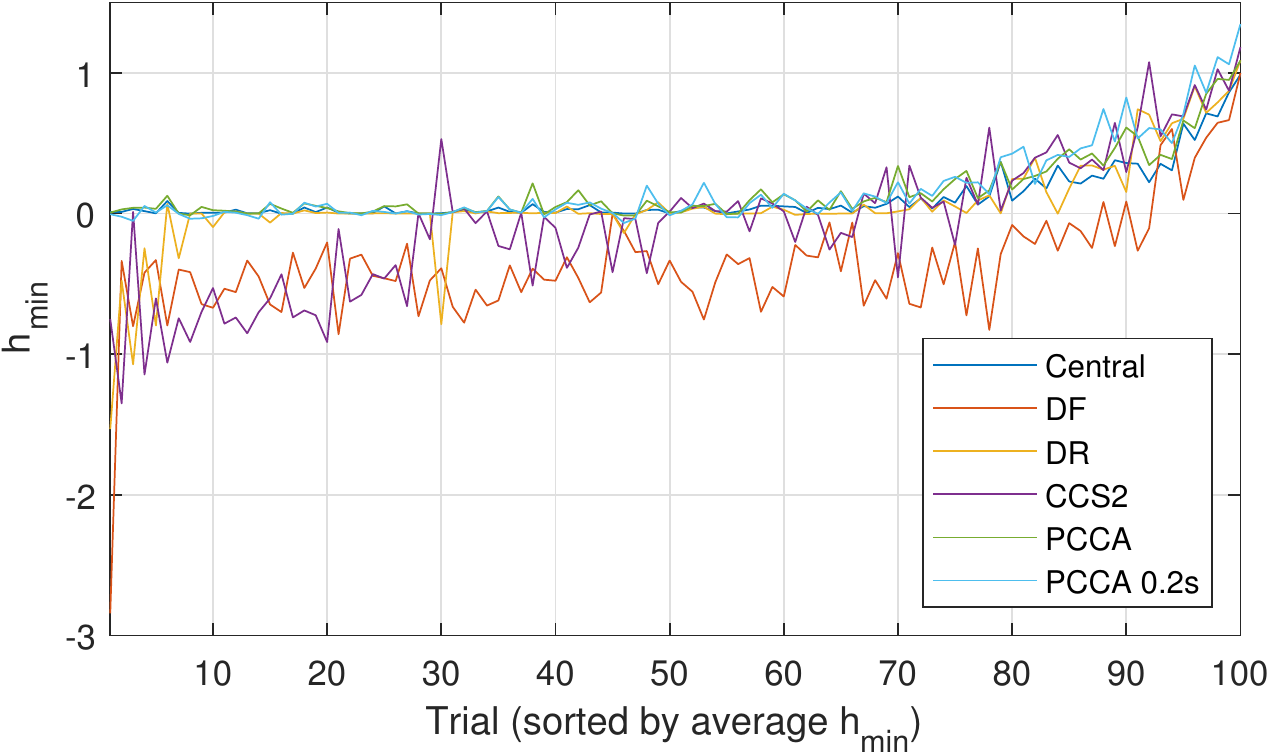}
    \vspace{-0.1in}
    \caption{Minimum barrier distance for 100 5-agent Monte Carlo simulations, sorted by average from min to max.}
    \vspace{-0.1in}
    \label{fig:hmin}
\end{figure}

For collision avoidance in the multi-agent case, the Centralized controller exhibits the best results, but also requires explicit communication. While the barrier is shown to be violated (i.e. $h_{min} = -0.002$, $h_{min}$ is the minimum agent-to-agent barrier value for all the agents during each run), this is due to the selection of sampling time. When the sampling time was reduced, this barrier violation disappeared as expected. Both the PCCA controller with unit delay as well as the PCCA controller with low-pass filter perform almost as good as the Centralized controller.  A pictorial comparison is shown in Figure \ref{fig:hmin} that displays the minimum agent-to-agent barrier value over each trial simulation for all methods, sorted by average  of $h_{min}$ over the algorithms. While DF and CCS2 generally have larger violations than the other methods, DR has only a few visible violations. Both the Centralized controller and PCCA controllers exhibit minimal barrier violation throughout all simulation trials and are almost indistinguishable in the plot.
%

We now rerun the Monte-Carlo simulation trials using the worst-case agent-to-agent barrier violations $h_{min}$ recorded in Table \ref{tab:compare} to add a radius margin ($r > 2 r_0$) for each algorithm's computation of the barrier constraint $h$. The results shown in Table \ref{tab:compare_marg} depict the barrier violations with the agents' actual size $r_0$; all methods but one CCS2 trial effectively avoid collision with radius margin added. The CCS2 trial with collision is due to an interation between agent-to-agent hard constraints and agent-to-static-outer-circle soft constraints. It is also noteworthy that the radius margins added for both DF and DR grow the agent sizes enough that there are additional no-converge as well as infeasible trials, compared to the zero-margin cases. One could iterate on the barrier margin required for both DF and DR to achieve $h_{min}$ closer to zero, similar to the other methods.
%
\begin{table}[b!]\vspace{-0.1in}
	\caption{Metrics for CBF-based collision-avoidance algorithms from 100 Monte-Carlo runs after adding worst-case agent-to-agent violation to radius margin.}	
	\begin{tabular}{l|ccc|c|c|c|}
	\cline{2-7}
	                                & \multicolumn{3}{c|}{Converge Time (sec)}             &                & \# no    & \#         \\
	                                & min                   & max                   & mean & $h_{\rm{min}}$ & converge & infeasible \\ \hline
	\multicolumn{1}{|l|}{Central}      & \multicolumn{1}{c|}{7.45} & \multicolumn{1}{c|}{22.15} & 12.98    & 0.000    & 0         & 0          \\ \hline
	\multicolumn{1}{|l|}{DF}           & \multicolumn{1}{c|}{7.90} & \multicolumn{1}{c|}{50.60} & 18.03    & 1.34      & 11         & 37        \\ \hline
	\multicolumn{1}{|l|}{DR}           & \multicolumn{1}{c|}{7.95} & \multicolumn{1}{c|}{82.95} & 19.91    & 1.67      & 5         & 38        \\ \hline
	\multicolumn{1}{|l|}{CCS2}         & \multicolumn{1}{c|}{7.80} & \multicolumn{1}{c|}{36.75} & 15.37    & -0.91      & 5         & 0          \\ \hline
	\multicolumn{1}{|l|}{PCCA}         & \multicolumn{1}{c|}{7.35} & \multicolumn{1}{c|}{23.75} & 12.77    & -0.002      & 0         & 0          \\ \hline
	\multicolumn{1}{|l|}{PCCA$_{0.2}$} & \multicolumn{1}{c|}{7.35} & \multicolumn{1}{c|}{21.80} & 12.70    & 0.001    & 0         & 0          \\ \hline
	\end{tabular}
	\vspace{-0.1in}
	\label{tab:compare_marg}
\end{table}
%

In the two-agent case, most of the methods reviewed in this paper work adequately well. In the absence of explicit communication in a multi-agent scenario, the PCCA controller exhibits the best overall performance, taking into account liveness, controller feasibility, and the required radius margin to avoid collisions. While a measurement, or estimate, of the other agents' accelerations is needed in the update of \eqref{hatwj}, this can generally be obtained from velocity measurements or estimates and appropriate differentiation or lead filtering, or with the use of an estimator (e.g. \cite{ansari}).

\section{Conclusions}
This paper compared several CBF-based algorithms for their performance in multi-agent scenarios. The results show
that algorithms taking all the constraints into account have lower convergence times and, as proven
in this paper, were always feasible. Adherence to constraints is more complex. The Centralized and PCCA
policies showed minimal violations while the Decentralized Follower and Reciprocal methods had a few larger violations due
to infeasibility.  In the latter case, the problem could be solved by braking, while in all cases, a
radius margin could be increased, both actions potentially negatively impacting liveness.  
\noindent


\end{document}